\documentclass{article}
\usepackage[backend=biber,style=ieee,maxnames=1,minnames=1]{biblatex}
\usepackage{spconf,amsmath,graphicx,hyperref}
\usepackage{booktabs,multirow,makecell}
\usepackage{kotex}
\usepackage[para]{footmisc}
\usepackage{xspace}
\usepackage[table]{xcolor}
\usepackage[most]{tcolorbox}
\usepackage[capitalise]{cleveref}
\usepackage{enumitem}
\definecolor{light_gray}{rgb}{0.85, 0.85, 0.85}

\usepackage{subcaption}

\newcommand{\benchmark}{\textsc{TwinShift}\xspace}

\newcommand{\tparagraph}[1]{\noindent\textbf{#1}}

\newcommand{\asvspoof}{\textit{ASVspoof'19}\xspace}

\newcommand{\astfootnote}[1]{
    \let\oldthefootnote=\thefootnote
    \setcounter{footnote}{1}
    \renewcommand{\thefootnote}{\fnsymbol{footnote}}
    \footnotetext{#1}
    \let\thefootnote=\oldthefootnote
}

\title{TwinShift: Benchmarking Audio Deepfake Detection across Synthesizer and Speaker Shifts}

\name{
\begin{tabular}{c}
Jiyoung Hong$^{1,\dagger}$ \quad Yoonseo Chung$^{1,\dagger}$ \quad Seungyeon Oh$^1$ \quad Juntae Kim$^2$ \\
\textit{Jiyoung Lee}$^{1,\star}$ \quad \textit{Sookyung Kim}$^{1,\star}$ \quad \textit{Hyunsoo Cho}$^{1,\star}$
\end{tabular}
\thanks{$^{\dagger}$ Equal contribution \\ $^{\star}$ Corresponding authors}
}

\address{
$^1$ Ewha Womans University \\
$^2$ SK Telecom, Seoul, Repulic of Korea
}

\begin{document}
\ninept
\maketitle

\begin{abstract}

     Audio deepfakes pose a growing threat, already exploited in fraud and misinformation. A key challenge is ensuring detectors remain robust to unseen synthesis methods and diverse speakers, since generation techniques evolve quickly. Despite strong benchmark results, current systems struggle to generalize to new conditions limiting real-world reliability.
     To address this, we introduce \benchmark, a benchmark explicitly designed to evaluate detection robustness under strictly unseen conditions. 
     Our benchmark is constructed from six different synthesis systems, each paired with disjoint sets of speakers, allowing for a rigorous assessment of how well detectors generalize when both the generative model and the speaker identity change. 
     Through extensive experiments, we show that \benchmark\ reveals important robustness gaps, uncover overlooked limitations, and provide principled guidance for developing ADD systems. The \benchmark\ benchmark
can be accessed at \url{https://github.com/intheMeantime/TWINSHIFT}.

\end{abstract}
\begin{keywords}
Audio deepfake, Benchmark, Generalization
\end{keywords}
\section{Introduction}
\label{sec:intro_en}
    
    The rapid advancement of neural speech synthesis technologies has brought both remarkable progress and alarming risks, where in the time it takes to watch a short clip, a stranger can now clone a voice that is nearly indistinguishable from the original.~\cite{rosi2025perception}
    While these breakthroughs enable beneficial applications such as personalized digital assistants~\cite{peek_digital_assistants_improve_workplace_productivity_2024} and accessibility for visually impaired users ~\cite{iflytek_tts_daily_life_2024, shekhawat_voice_assistants_education_2023} they also open the door to malicious misuse. 
    Specifically, audio deepfakes have already been exploited in high-profile fraud cases~\cite{stupp2019fraud}, political misinformation campaigns ~\cite{meaker_wired_deepfake_audio_political_nightmare_2023}, and social engineering attacks~\cite{Hutiri_2024}, amplifying concerns about public trust and security.~\cite{detectionsurvey}
    These growing risks have placed Audio Deepfake Detection (ADD) \cite{Yamagishi2019, 2021asvspoof, wang2024asvspoof5crowdsourcedspeech}  at the center of defense efforts, raising a critical question: not whether detectors can recognize \textit{yesterday’s fakes}, but whether a detector trained today will still catch \textit{tomorrow’s voices}.

    Yet current ADD systems often memorize spurious cues tied to specific generators ~\cite{pianese2022deepfakeaudiodetectionspeaker, kheir2025viewstruthspectralselfsupervised}, speakers, or preprocessing pipelines; when a new synthesis method appears, these cues disappear and performance collapses.~\cite{nguyen2025read, anomalydetectionlocalizationspeech} Robust out-of-distribution (OOD) generalization is therefore a deployment requirement. 
    To keep pace with new synthesis models, most studies have adopted a reactive strategy: folding outputs from each newly released model into a growing composite dataset and then applying random train–test splits\cite{huang2025speechfake}. 
    While such setups follow common ML practice, this practice leaks distributional information and systematically overestimates robustness—detectors look strong on in-distribution benchmarks yet fail on truly novel methods \cite{ITW}.
    This gap is not just hypothetical. Recent studies \cite{bhagtani2024diffssddiffusionbaseddatasetspeech,Bhagtani2024RecentDeepfake, huang2025speechfake} have confirmed it by showing striking performance drops when ADD models are evaluated under controlled OOD settings, such as detecting voices from unseen speakers or audio produced by entirely new synthesis pipelines \cite{ITW, Bhagtani2024RecentDeepfake}.
    Together, these findings highlight a growing mismatch between current evaluation methodology and the demands of real-world deployment.

    To address these limitations, we propose \benchmark, a new benchmark explicitly designed to measure and stress-test the generalization ability of ADD systems. Our contributions are as follows:
\begin{itemize}[leftmargin=*]
    
    \item \textbf{OOD Composite Benchmarking.} We construct a dataset where evaluation is strictly performed on unseen conditions. Specifically, our test data contains audio from unseen (i) speakers (ii) synthesizer. The benchmark is built using six synthesizer, with distinct speakers across models, ensuring that evaluation samples are disjoint from training data at both the speaker and synthesizer.
    
    \item \textbf{Empirical Evaluation Across SOTA Models}. We conduct extensive experiments with a wide range of state-of-the-art synthesis models to validate the benchmark. Our analysis diagnoses current limitations and outlines pathways toward ADD resilience to the rapidly evolving synthesis landscape.
    
\end{itemize}

\section{Preliminary Study}
\label{sec:02analysis}

\begin{table}[t!]
    \centering
    \caption{Preliminary results. Training and seen-model evaluation use F5-TTS–generated spoofs; unseen-model evaluation uses HierSpeech++. All sets contain 50 speakers per class.}
    
\resizebox{1.0\columnwidth}{!}{
{\renewcommand{\arraystretch}{1.1}
    \begin{tabular}{c|cccc}
    \toprule
    \multirow{2}{*}{\makecell{\textbf{Detection} \\ \textbf{Model}} } 
        & \multicolumn{2}{c}{\textbf{Seen Model}} 
        & \multicolumn{2}{c}{\textbf{Unseen Model}} \\
    \cmidrule(lr){2-5} 
        & \textbf{Seen Spk.} & \textbf{Unseen Spk.} 
        & \textbf{Seen Spk.} & \textbf{Unseen Spk.} \\

    \midrule
    Se-Res2Net & 0.014 & 0.074 & 0.394 & 0.578 \\
    RawNet2 & 0.002 & 0.066 & 0.326 & 0.466 \\
    AASIST & 0.016 & 0.060 & 0.390 & 0.588 \\
    RawBmamba & 0.006 & 0.006 & 0.472 & 0.560 \\

    \midrule
    Average & 0.010 & 0.052 & 0.396 & 0.548 \\ 

    \bottomrule
\end{tabular}}}

    \label{tab:pre_anal}
\end{table}
% \vspace{-8pt}

\begin{table*}[ht]
    \centering
    \caption{Description of \benchmark. ASV refers to \textit{ASVspoof 2019 LA train}, and ITW refers to \textit{In-the-Wild}. The ratios of bonafide and spoof samples in utt\_train and utt\_test are consistent with their proportions in the total number of utterances.}
    \resizebox{\textwidth}{!}{
\renewcommand{\arraystretch}{1.1} 
\setlength{\tabcolsep}{9pt}
\begin{tabular}{ccccccccc}
    \toprule
    % \textbf{ID} & \textbf{Bonafide} & \textbf{Spoof} & \textbf{Types} & \textbf{\# Speakers} & \textbf{Duration} & \textbf{\# Utterances} & \textbf{\# utt\_train} & \textbf{\# utt\_test} \\
    % & & & & \textbf{(bona, spoof)} & \textbf{(hours)} & \textbf{(bona, spoof)} & & \\
    \textbf{ID} & \textbf{Bonafide} & \textbf{Spoof}
    & \makecell{\textbf{Generator} \\ \textbf{Type}} 
    & \makecell{\textbf{\# Speakers} \\ \textbf{(bona, spoof)}} 
    & \makecell{\textbf{Duration} \\ \textbf{(hours)}} 
    & \makecell{\textbf{\# Utterances} \\ \textbf{(bona, spoof)}} 
    & \textbf{\# utt\_train} & \textbf{\# utt\_test} \\
    
    \midrule
    Mai & ASV, ITW & MeloTTS       & TTS   & (24, 5)       & 77.35 & 39059 (3911,35184)  & 31680 & 7415 \\
    Pai & ASV, ITW & ParlerTTS     & TTS   & (24, 5)       & 67.07 & 35297 (3911,31386)  & 28314 & 7063 \\
    Eai & ASV, ITW & ElevenLabs    & VC    & (23, 20)      & 5.02  & 6071 (668,5403)     & 4866  & 1205 \\
    Hex & Expresso & Hierspeech++  & zsTTS & (4, 4)        & 9.82  & 11000 (1000,10000)  & 8801  & 2199 \\
    Fem & Emilia   & F5-TTS        & zsTTS & (365, 365)    & 18.21 & 11198 (1198,10000)  & 8843  & 2281 \\
    Oli & LibriTTS & OZspeech      & zsTTS & (123, 123)    & 14.39 & 11080 (1081,9999)   & 8971  & 2109 \\
    \bottomrule
\end{tabular}}
%\vspace{-8pt}
    \label{tab:dataset}
\end{table*}

Before presenting our benchmark, we emphasize two defining axes of its design:\textbf{(i) synthesis model} and \textbf{(ii) speaker identity}. 
These factors encapsulate the essential risk of encountering unseen voices or generators in the real-world while remaining experimentally tractable. 
The following sections detail how ADD systems generalize to unseen speakers or audio generators individually, and how simultaneous shifts along both axes compound these challenges.

\subsection{Experimental Setup}
To examine both localized transferability along the two axes, \emph{speaker identity} and \emph{synthesis model}, and their compounded effect when combined, we conduct a controlled study that partitions each axis into \emph{seen} and \emph{unseen} conditions.
 
\tparagraph{Axis 1 (Synthesis model):} 
We select 2 different synthesis models when generating the spoof dataset: \emph{HierSpeech++}~\cite{lee2023hierspeechbridginggapsemantic} and \emph{F5-TTS}~\cite{chen-etal-2024-f5tts}.
These models reflect fundamentally different paradigms, with HierSpeech++ based on hierarchical latent factorization \cite{lee2023hiervsthierarchicaladaptivezeroshot} and F5-TTS on flow-matching transport \cite{lipman2023flowmatchinggenerativemodeling}.
Importantly, both can generate speech from speakers unseen during training, enabling precise control of speaker visibility and directly supporting our second axis, \emph{speaker identity}.
In our setup, the detector is trained on spoof audio synthesized with F5-TTS to establish a consistent training distribution, while evaluation includes both F5-TTS and HierSpeech++ outputs, ensuring that we can measure within-generator generalization as well as transferability across fundamentally different synthesis models.

\tparagraph{Axis 2 (Speaker identity):} 
To guarantee that speaker identities remain strictly disjoint between train-test sets, we draw all speakers from the Emilia~\cite{he2024emiliaextensivemultilingualdiverse} bonafide dataset and partition them into two non-overlapping groups.
We sample 100 distinct speakers, designating 50 as the \emph{seen} pool and using their real speech as bonafide; the remaining 50 are held out and used only to synthesize zero-shot spoof samples with \emph{each} generator (F5-TTS and HierSpeech++).

\tparagraph{Evaluation condition:} 
Building on the two design axes, we evaluate ADD systems under a comprehensive set of controlled conditions to isolate the impact of each factor and to examine their compounded effect.
Each condition specifies whether the synthesis model and speaker identities are \emph{seen} or \emph{unseen} relative to training as follows:

\begin{itemize}[leftmargin=*]
    \item \textbf{Seen Model / Seen Speaker:} Both the synthesis model and speaker identities overlap with training, serving as a baseline.
    \item \textbf{Seen Model / Unseen Speaker:} Spoof audio is synthesized with the same model as training, but evaluation speakers are disjoint, isolating generalization across speakers.
    \item \textbf{Unseen Model / Seen Speaker:} Evaluation uses spoof audio synthesized with a different generator (HierSpeech++) while speaker identities overlap, leaving transferability across synthesis models.
    \item \textbf{Unseen Model / Unseen Speaker:} Both the generator and speakers are unseen, producing the most challenging cross-axis shift.
    % \item \textbf{Asymmetric Speaker Conditions:} To further disentangle the role of familiarity, we include cases where only bonafide speakers are seen (spoof unseen) or only spoof speakers are seen (bonafide unseen).
\end{itemize}

\tparagraph{Detection models \& Evaluation metric:} 
For evaluation, we selected four widely used audio deepfake detectors: \emph{Se-Res2Net}~\cite{Gao_2021}, \emph{RawNet2}~\cite{rawnet2}, \emph{AASIST}~\cite{aasist}, and \emph{RawBMamba}~\cite{rawbmamba}. 
These models span different design approaches, giving us a broad view of detector behavior across architectures. 
Detailed descriptions of these detectors are provided in Sec.~\ref{ssec:detectors}.
As the evaluation metric, we report \emph{Equal Error Rate (EER)}, which balances false accept and false reject rates and has been the standard measure in audio spoofing benchmarks~\cite{2021asvspoof} .

\subsection{Results}
Table~\ref{tab:pre_anal} presents the outcomes of our preliminary study.  
When both axes are aligned with training (Seen Model / Seen Speaker), detectors perform reliably, with a macro-average EER of only ($\approx$ 0.010).  
    
\tparagraph{Single-axis shifts:}
Varying one factor at a time reveals that both axes are important, but to different degrees. 
Changing only speaker identity under a seen generator increases the average EER from \(\approx 0.010\) to \(\approx 0.052\) (absolute \(+0.042\)).  
By contrast, changing only the synthesis model while keeping speakers seen raises the error to \(\approx 0.396\) (absolute \(+0.386\)).  
Thus, generator mismatch is the dominant source of degradation, while speaker changes under a fixed generator have a smaller effect.

\tparagraph{Combined shifts:}
When both axes change simultaneously (model / speaker), the average EER reaches \(\approx 0.548\).  
Relative to the model-only shift (\(\approx 0.396\)), introducing unseen speakers adds a further \(+0.152\) EER.  
This shows that speaker identity \emph{modulates} difficulty, with its impact most visible once the generator itself has shifted.

\tparagraph{Takeaways:}
Taken together, these findings indicate that the two axes exert asymmetric but complementary effects.  
Speaker identity alone induces only modest degradation, whereas synthesis-model shifts account for the majority of errors.  
However, when both factors are simultaneously unseen, their effects accumulate, producing the most challenging condition, which may reflect a compounding distribution shift across the two axes.

\begin{table*}[ht]
    \centering
    \caption{EER results on our dataset, which comprises six disjoint environments. The diagonal entries correspond to cases where both the synthesizer and speaker are seen, while all other entries represent cases where both axes are unseen. For each model, the best-performing test set is indicated in \textbf{bold}, and the second-best in \underline{underline}.}
    \resizebox{\textwidth}{!}{
\setlength{\tabcolsep}{8pt}
\begin{tabular}{l|c | c c c c c c | c|c}
    \toprule
     % & \textbf{Detection} & \textbf{dataset A} & \textbf{dataset B} & \textbf{dataset C} 
     %  & \textbf{dataset D} & \textbf{dataset E} & \textbf{dataset F} & \textbf{Unseen}\\
     % & \textbf{Model} & (melo) & (parler) & (elevenlabs) & (hier) & (F5tts) & (oz) &  \textbf{Avg. EER}\\
    & Method 
    & $Mai_{test}$ & $Pai_{test}$ & $Eai_{test}$ & $Hex_{test}$ & $Fem_{test}$ & $Oli_{test}$
    & \makecell{Unseen \\ Avg. EER}
    & \makecell{All \\ Unseen}\\
    \midrule
    \multirow{4}{*}{$Mai_{train}$}
        & Se-Res2Net & \cellcolor{light_gray}\textbf{0.00491} & \underline{0.03315} & 0.07257 & 0.45011 & 0.40283 & 0.83713 & 0.35916 & 0.24421 \\
        & RawNet2    & \cellcolor{light_gray}\textbf{0.00266} & \underline{0.01542} & 0.03990 & 0.42010 & 0.32878 & 0.74961 & 0.31076 & 0.23481 \\
        & AASIST     & \cellcolor{light_gray}\textbf{0.00316} & \underline{0.03807} & 0.03990 & 0.50987 & 0.49827 & 0.76428 & 0.37008 & 0.25610 \\
        & RawBmamba  & \cellcolor{light_gray}\underline{0.00267} & 0.00890 & \textbf{0.00046} & 0.45536 & 0.34562 & 0.79666 & 0.32140 & 0.22034\\
    % \cmidrule(lr){2-9}
    %     & Avg.  & 0.00335 & 0.02389 & 0.03821 & 0.45886 & 0.39388 & 0.78692 & - \\
    \midrule
    \multirow{4}{*}{$Pai_{train}$}
        & Se-Res2Net & \underline{0.01052} & \cellcolor{light_gray}\textbf{0.00143} & 0.02404 & 0.52988 & 0.33949 & 0.73116 & 0.32702 & 0.23218 \\
        & RawNet2    & \underline{0.03682} & \cellcolor{light_gray}\textbf{0.00254} & 0.16780 & 0.46536 & 0.38626 & 0.68285 & 0.34782 & 0.25836 \\
        & AASIST     & 0.11265 & \cellcolor{light_gray}\underline{0.01017} & \textbf{0.00816} & 0.48562 & 0.45201 & 0.45201 & 0.30209 & 0.31505 \\
        & RawBmamba  & 0.05002 & \cellcolor{light_gray}\textbf{0.00016} & \underline{0.00093} & 0.50613 & 0.30687 & 0.78857 & 0.33050 & 0.22716 \\
    % \cmidrule(lr){2-9}
    %      & Avg.  & 0.05250 & 0.00358 & 0.05023 & 0.49675 & 0.37116 & 0.66365 & - \\
    \midrule
    \multirow{4}{*}{$Eai_{train}$}
        & Se-Res2Net & \underline{0.02631} & 0.04937 & \cellcolor{light_gray}\textbf{0.00093} & 0.50413 & 0.45763 & 0.72357 & 0.35220 & 0.19510 \\
        & RawNet2    & 0.09800 & \underline{0.04030} & \cellcolor{light_gray}\textbf{0.00000} & 0.50037 & 0.46350 & 0.65857 & 0.35215 & 0.21551 \\
        & AASIST     & 0.14472 & \underline{0.01661} & \cellcolor{light_gray}\textbf{0.00816} & 0.51412 & 0.45710 & 0.66742 & 0.35999 & 0.23600 \\
        & RawBmamba  & \underline{0.00526} & 0.00644 & \cellcolor{light_gray}\textbf{0.00000} & 0.57489 & 0.22454 & 0.69753 & 0.30173 & 0.15247 \\
    % \cmidrule(lr){2-9}
    %     & Avg. & 0.06857 & 0.02818 & 0.00227 & 0.52338& 0.40069 & 0.68677 & - \\
    \midrule
    \multirow{4}{*}{$Hex_{train}$}
        & Se-Res2Net & 0.32894 & \underline{0.20049} & 0.31387 & \cellcolor{light_gray}\textbf{0.02376} & 0.50147 & 0.52099 & 0.37315 & 0.34699  \\
        & RawNet2    & 0.11055 & 0.14595 & \underline{0.10385} & \cellcolor{light_gray}\textbf{0.01025} & 0.48438 & 0.48610 & 0.26616 & 0.20387 \\
        & AASIST     & 0.24209 & \underline{0.13450} & 0.25577 & \cellcolor{light_gray}\textbf{0.03525} & 0.52338 & 0.65857 & 0.36286 & 0.26118 \\
        & RawBmamba  & 0.37103 & 0.31219 & \underline{0.18274} & \cellcolor{light_gray}\textbf{0.00050} & 0.49318 & 0.54476 & 0.38078 & 0.37339 \\
    % \cmidrule(lr){2-9}
    %     & Avg. & 0.26315 & 0.19828 & 0.21406 & 0.01744 & 0.50060 & 0.55260 & - \\
    \midrule
    \multirow{4}{*}{$Fem_{train}$}
        & Se-Res2Net & 0.42631 & 0.58121 & \underline{0.22404} & 0.52488 & \cellcolor{light_gray}\textbf{0.04704} & 0.54476 & 0.46024 & 0.49268 \\
        & RawNet2    & 0.39733 & \underline{0.29565} & 0.40724 & 0.55588 & \cellcolor{light_gray}\textbf{0.00534} & 0.45649 & 0.42252 & 0.37125 \\
        & AASIST     & 0.41837 & 0.40098 & \underline{0.34422} & 0.54488 & \cellcolor{light_gray}\textbf{0.03848} & 0.53667 & 0.44902 & 0.43440 \\
        & RawBmamba  & 0.53441 & 0.57485 & \underline{0.07211} & 0.75994 & \cellcolor{light_gray}\textbf{0.02192} & 0.62620 & 0.51350 & 0.52922 \\
    % \cmidrule(lr){2-9}
    %     & Avg. & 0.44410 & 0.46317 & 0.26190 & 0.59640 & 0.02819 & 0.54103 & - \\
    \midrule
    \multirow{4}{*}{$Oli_{train}$}
        & Se-Res2Net & 0.96317 & 0.80722 & 0.88891 & \underline{0.51913} & 0.57577 & \cellcolor{light_gray}\textbf{0.00885} & 0.75084 & 0.77340 \\
        & RawNet2    & 0.94170 & 0.87328 & 0.89614 & 0.58515 & \underline{0.53917} & \cellcolor{light_gray}\textbf{0.00151} & 0.63949 & 0.83262 \\
        & AASIST     & 0.96576 & 0.80459 & 0.83359 & \underline{0.50962} & 0.57818 & \cellcolor{light_gray}\textbf{0.01467} & 0.73835 & 0.76050 \\
        & RawBmamba  & 0.96577 & 0.93012 & 0.98413 & 0.59115 & \underline{0.57257} & \cellcolor{light_gray}\textbf{0.00809} & 0.80875 & 0.81431 \\
    % \cmidrule(lr){2-9}
    %     & Avg. & 0.95910 & 0.85380 & 0.90069 & 0.55126 & 0.56642 & 0.00828 & - \\
    % \midrule
    % \multirow{4}{*}{\makecell[l]{dataset All \\ {A-F}}}
    %     & Se-Res2Net & 0.00778 & 0.00143 & 0.01587 & 0.03900 & 0.03020 & 0.02251 & - & - \\
    %     & RawNet2    & 0.00309 & 0.00254 & 0.00677 & 0.02000 & 0.02459 & 0.01618 & - & - \\
    %     & AASIST     & 0.01318 & 0.00755 & 0.00816 & 0.10502 & 0.04651 & 0.01694 & - & - \\
    %     & RawBmamba  & 0.00267 & 0.00254 & 0.00817 & 0.00400 & 0.00320 & 0.00835 & - & - \\

    \bottomrule
\end{tabular}}
    \label{tab:mainresult}
\end{table*}

\section{\benchmark Benchmark}

\label{sec:03method}
Expanding on the preliminary study, we introduce \textsc{TwinShift}, a benchmark \emph{structured around two orthogonal axes}—synthesis model and speaker identity—with explicit seen/unseen visibility controls. 
Unlike prior approaches~\cite{huang2025speechfake,kumar2025indiefakedatasetbenchmarkdataset} that rely on a single pooled split, we construct \textbf{six mutually disjoint environments}, each pairing one-to-one dedicated \emph{bonafide} dataset with one spoofing system.  
\benchmark\ supports within-environment baselines and cross-environment transfer, yielding a more faithful view of real-world robustness.

\tparagraph{Bonafide Sources:}
\label{ssec:bonafide}
Bonafide utterances are drawn from five widely used corpora: \asvspoof LA train~\cite{Yamagishi2019}, In-the-Wild~\cite{ITW}, Expresso~\cite{nguyen2023expressobenchmarkanalysisdiscrete}, Emilia~\cite{he2024emiliaextensivemultilingualdiverse}, and LibriTTS train-clean-100~\cite{zen2019librittscorpusderivedlibrispeech}.  
To prevent leakage across conditions, each corpus is assigned to one environment in the benchmark, forming a self-contained bonafide–spoof pair, excluding \asvspoof LA train and In-the-Wild which are split across three environments, but still with disjoint speaker partitions to ensure no overlap.

\tparagraph{Spoof Generation:}
\label{ssec:spoof}
To synthesize spoofed audio, we employ six representative TTS and voice-conversion systems spanning diverse generative paradigms: MeloTTS~\cite{zhao2024melo}, HierSpeech++~\cite{lee2023hierspeechbridginggapsemantic}, ParlerTTS~\cite{lyth2024natural}, F5-TTS~\cite{chen-etal-2024-f5tts}, OZSpeech~\cite{huynhnguyen2025ozspeechonestepzeroshotspeech}, and the ElevenLabs API~\cite{ElevenLabs2023}.
These systems differ in architecture and conditioning mechanisms, ensuring broad coverage across model-level variability.  Moreover, these models leverage two distinct conditioning strategies: (i) \textbf{predefined speakers} (MeloTTS, ParlerTTS, ElevenLabs), where the model comes with a built-in set of independent speakers; and (ii) \textbf{zero-shot speakers} (HierSpeech++, F5-TTS, OZSpeech), where spoof samples are generated from reference utterances in the paired bonafide, producing unseen speakers by construction.  

\tparagraph{Dataset Composition and Splitting:}
\label{ssec:composition}
The overall composition of \benchmark\ is summarized in Table~\ref{tab:dataset}.
Each bonafide corpus is paired with a spoofing system based on two principles: (i) \textbf{experimental coherence}, aligning corpora with generators they are trained on or closely associated with (e.g., Emilia with F5-TTS, LibriTTS with OZSpeech, Expresso with HierSpeech++), and (ii) \textbf{speaker disjointness}, ensuring no speaker overlaps across environments.
For more general corpora such as ASVspoof and ITW, speakers are partitioned and assigned to different environments without overlap.
Each environment (Mai, Pai, Eai, Hex, Fem, Oli) contains paired bonafide and spoof subsets.  
Following the ASVspoof convention~\cite{2019asvspoof}, we maintain a $1{:}9$ class ratio and then split each environment $8{:}2$ into training and evaluation partitions.  
Within this protocol, we instantiate: \emph{(i) a fully seen baseline} (model and speakers seen), and \emph{(ii) the combined-shift condition} (model and speakers both unseen) as the \textbf{primary} evaluation target, motivated by the preliminary findings. 
This setup mirrors the preliminary design (Sec.~\ref{sec:02analysis}) while scaling it across multiple, disjoint environments, thereby enabling controlled stress-tests of generalization in both axes simultaneously.

\section{Experiments}
\label{sec:04exp}

% We use \benchmark\ to evaluate whether current audio deepfake detectors remain reliable when synthesizer and speaker identity shifts simultaneously.  

\subsection{Detectors and Training Protocol}
\label{ssec:detectors}
To obtain a broad and representative view, we evaluate four detectors drawn from distinct architectural families for our benchmark:

% HS
\begin{itemize}[leftmargin=*]
    \item \textbf{Se-Res2Net ~\cite{Gao_2021}:} It is constructed by combining residual connections with multi-scale convolutions, while incorporating Squeeze-and-Excitation (SE) blocks to recalibrate channel importance, thereby enabling precise extraction of spectral features from speech signals.
    \item \textbf{RawNet2 ~\cite{rawnet2}:} RawNet2 employs 1D convolutional blocks combined with a GRU-based sequence encoder to directly learn time-frequency patterns from raw audio inputs.
    \item \textbf{AASIST ~\cite{aasist}:} AASIST is a GNN-based architecture that leverages a heterogeneous stacking Graph Attention Network to model temporal and spectral node-level features, while multi-head attention is applied to effectively capture spoofing artifacts.
    \item \textbf{RawBMamba ~\cite{rawbmamba}:} Built upon the Mamba framework, RawBmamba integrates bidirectional state-space blocks with multi-scale convolutions, enabling the simultaneous modeling of both short-term and long-term dependencies in audio signals.
\end{itemize}

\noindent Each detector is trained within a single environment $E_i$ (Mai, Pai, Eai, Hex, Fem, Oli) and then evaluated across all environments to assess both in-domain performance and cross-environment transferability using EER as a metric.
Table~\ref{tab:mainresult} reports the results, with diagonal entries ($E_i \!\rightarrow\! E_i$) as fully-seen in-domain baselines and off-diagonal entries ($E_i \!\rightarrow\! E_{j(i\neq j)}$) as cross-environment evaluations.

\subsection{Main Results}
\label{ssec:main}
As can be seen from Table.~\ref{tab:mainresult} two consistent patterns emerge:

\tparagraph{(1) Reliable in-domain, fragile cross-environment:}
Within a single environment, detectors achieve near-perfect accuracy, often with EERs approaching zero for simpler cases such as Mai, Pai, and Eai.
Yet this reliability collapses the moment evaluation crosses into a different environment: scores degrade sharply, especially on more challenging generators, such as OZSpeech ($E_{Oli}$). 
These results confirm that the two axes—synthesizer and speaker identity—are pivotal determinants of robustness.

\tparagraph{(2) No single model / dataset stands resilient to shifts:}
Looking across environments, no detector architecture consistently outperforms the rest, and no training dataset provides immunity against shifts. 
RawNet2 shows relatively stronger portability, but it even fails on the hardest targets. 
High-fidelity generators such as F5-TTS ($E_{Fem}$) or OZSpeech ($E_{Oli}$), which one might expect to offer better coverage, instead exhibit limited generalizability across environments, indicative of overfitting to narrow artifacts. 
These findings suggest that resilience cannot be secured by `picking the right model' or `training on the right data' alone, indicating true progress requires advances on both fronts.

\section{Discussion}

\subsection{Transfer is Non-Commutative}
\label{ssec:noncommutative}

Another interesting finding is that transfer between environments is highly \emph{non-commutative}.  
That is, performance in direction $E_i \!\rightarrow\! E_j$ often differs substantially from $E_j \!\rightarrow\! E_i$.  
This skew indicates that artifacts left by different synthesizers are not symmetric or interchangeable, but instead manifest in generator-specific ways.
Several pairs illustrate this phenomenon clearly.  
For example, detectors trained on \textbf{Pai} transfer poorly to \textbf{Oli}, while the reverse direction is even more fragile, suggesting that the consistent artifacts of ParlerTTS do not prepare models for the bonafide-adjacent distribution of OZSpeech.  
A similar asymmetry appears between \textbf{Mai} and \textbf{Oli}, where learning from the simpler MeloTTS environment does not equip detectors for the harder OZSpeech target, but training on Oli also fails to generalize back to Mai due to overfitting to dataset-specific cues.  
We also observe asymmetric transfer within difficult pairs such as \textbf{Fem} and \textbf{Oli}, or \textbf{Hex} and \textbf{Fem}, where differences in distributional hardness yield mismatched transfer gains depending on the training direction.
These asymmetries likely arise because each synthesizer leaves qualitatively different traces: some produce consistent, salient artifacts that generalize outward, while others produce subtle or entangled artifacts that collapse when transferred.  
As a result, the transfer matrix is inherently skewed, and evaluating only one direction risks overlooking critical vulnerabilities.

\begin{figure}[t!]
    \begin{subfigure}{.48\linewidth}
        \centering
        \includegraphics[width=0.99\linewidth]{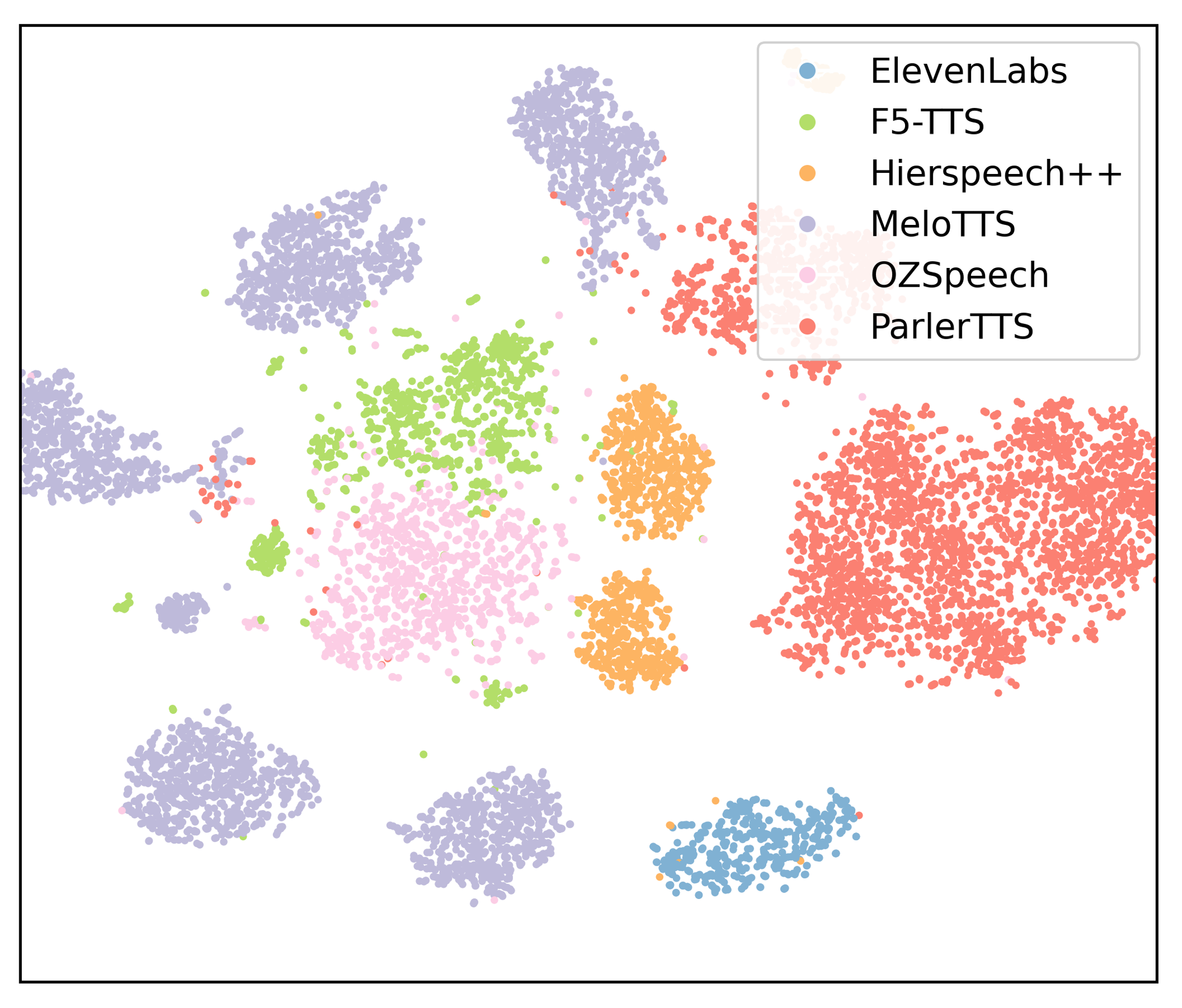}
        \label{fig:tsnesub1}
        \caption{Distribution of spoof audio}
    \end{subfigure}
    \begin{subfigure}{.48\linewidth}
        \centering
        \includegraphics[width=0.99\linewidth]{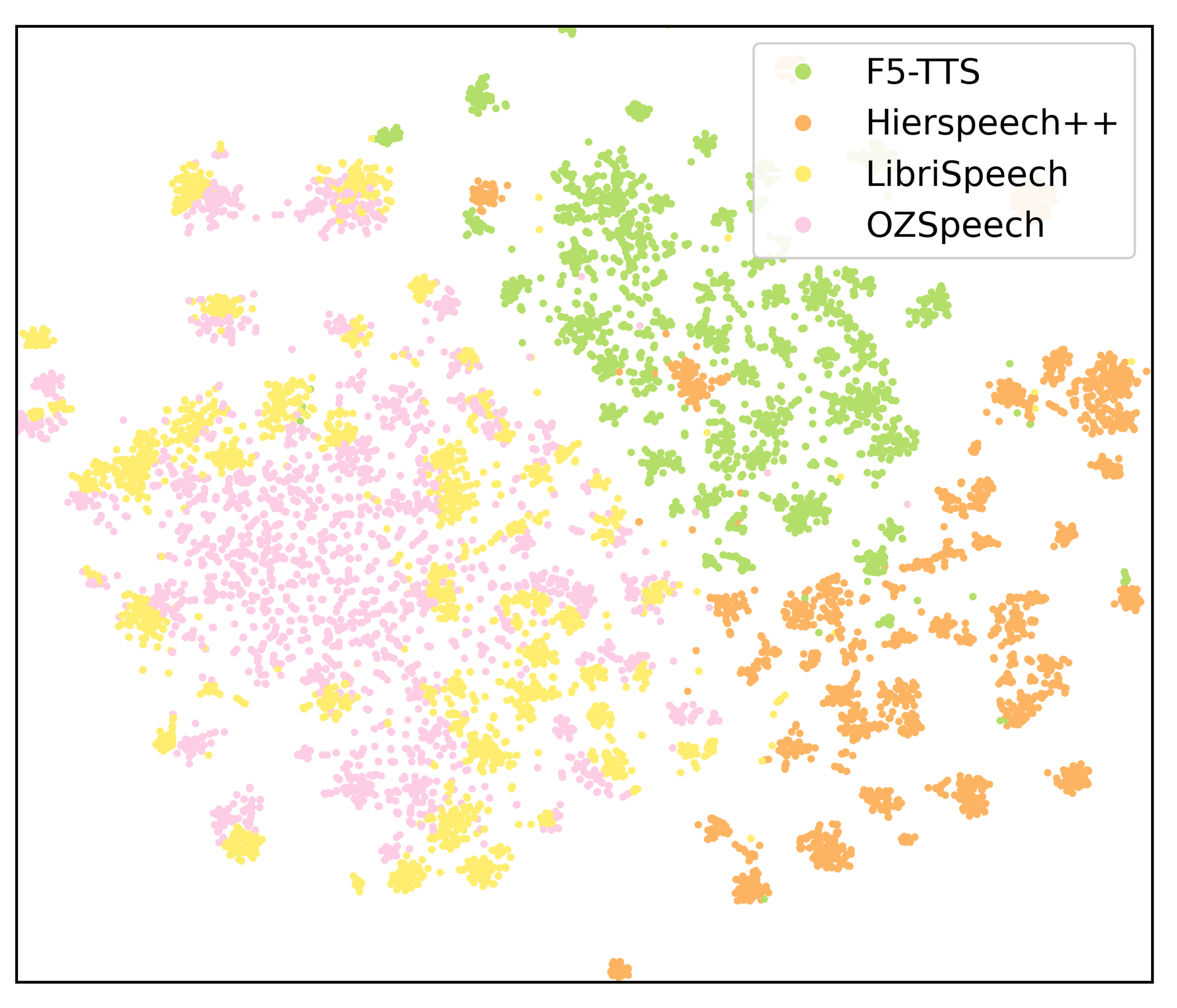}
        \label{fig:tsnesub2}
        \caption{Distribution of zero-shot TTS}
    \end{subfigure}

    \caption{t-SNE visualizations of audio data. (a) Spoof audio from \benchmark, showing the distribution of different spoofing methods. (b) Zero-shot TTS-generated speech, illustrating how each synthesizer’s output compares to bonafide speech.
    }
    \label{fig:tsne}
\end{figure}
\vspace{-8pt} 

\subsection{Bonafide is Composite; Spoofs Chase Islands}
\label{ssec:islands}

As shown in Fig.~\ref{fig:tsne}(b), bonafide utterances (yellow points) form multiple distinct clusters in the t-SNE space — appearing as a constellation of \emph{islands} rather than a single unified distribution. 
Among the spoof generators, OZSpeech in particular appears to \emph{chase} these islands, placing its embeddings close to certain bona fide clusters, yet failing to cover the entire landscape. 
This selective overlap illustrates why some regions of bonafide space are harder to defend against, while others remain relatively separable.
This mechanism helps explain both (i) why high-fidelity spoofs can be harder \emph{without} yielding broad transfer, and (ii) why transfer is non-commutative (\S\ref{ssec:noncommutative}): $E_i\!\to\!E_j$ is easier precisely when $E_i$ covers the islands emphasized by $E_j$, but the reverse need not hold.

\subsection{High-fidelity Spoofs Don’t Always Transfer}
\label{ssec:hiqual_hard}

A consistent pattern is that \emph{higher-quality} synthesizers (e.g., F5-TTS, OZSpeech) are \emph{harder to detect}, as their embeddings lie closer to bonafide regions in t-SNE (Fig.~\ref{fig:tsne}).
Crucially, however, training on such difficult, bonafide–like spoofs does \emph{not} guarantee broad transfer.
Models fit on high-fidelity generators learn subtle, generator-specific cues that fail to carry over to other environments, whereas training on sources with more \emph{consistent} artifacts (e.g., ParlerTTS) can yield stronger cross-environment performance.  
 Based on our observations, we discuss that robustness may depends on the \emph{combination} of (i) closeness to bonafide (task difficulty) and (ii) artifact consistency/diversity (transferability), not on fidelity alone. This appears to contrast with the common intuition that ``more realistic training data guarantees greater robustness''.
\section{Conclusion}
\label{sec:conclusion}

We introduced \textsc{TwinShift}, a benchmark that evaluates audio deepfake detectors under dual shifts of synthesizer and speaker identity.
Our results show that both factors can substantially degrade performance, transfer patterns between generators are highly asymmetric, and robustness cannot be secured by detector choice or training data alone.
By surfacing these challenges, \textsc{TwinShift} provides a foundation for building detectors capable of withstanding unseen and evolving spoofing attacks.

\clearpage

\printbibliography

\end{document}